\documentclass[aps,prl,reprint,preprintnumbers,superscriptaddress,amsmath,amssymb,bibnotes,longbibliography]{revtex4-1}

\usepackage{graphicx}
\usepackage{dcolumn}
\usepackage{bm}
\usepackage[colorlinks,linkcolor=blue,anchorcolor=blue,citecolor=blue,urlcolor=blue,filecolor=blue,menucolor=blue,runcolor=blue]{hyperref}

\begin{document}
\title{ Suppression of ferromagnetism and influence of disorder in silicon-substituted CeRh$_6$Ge$_4$ }
\author{Y. J. Zhang}
\affiliation{Institute for Advanced Materials, Hubei Normal University, Huangshi 435002, China
}
\affiliation{Center for Correlated Matter and Department of Physics,
Zhejiang University, Hangzhou 310058, China}
\author{Z. Y. Nie}
\affiliation{Center for Correlated Matter and Department of Physics,
Zhejiang University, Hangzhou 310058, China}
\author{R. Li}
\affiliation{Center for Correlated Matter and Department of Physics,
Zhejiang University, Hangzhou 310058, China}
\author{Y. C. Li}
\affiliation{Beijing Synchrotron Radiation Facility, Institute of High Energy Physics, Chinese Academy of Sciences, Beijing 100049, China}
\author{D. L. Yang}
\affiliation{Beijing Synchrotron Radiation Facility, Institute of High Energy Physics, Chinese Academy of Sciences, Beijing 100049, China}
\author{B. Shen}
\affiliation{Center for Correlated Matter and Department of Physics,
Zhejiang University, Hangzhou 310058, China}
\author{C. Ye}
\affiliation{Center for Correlated Matter and Department of Physics,
Zhejiang University, Hangzhou 310058, China}
\author{H. Su}
\affiliation{Center for Correlated Matter and Department of Physics,
Zhejiang University, Hangzhou 310058, China}
\author{R. Shi}
\affiliation{Institute for Advanced Materials, Hubei Normal University, Huangshi 435002, China
}
\author{S. Y. Wang}
\affiliation{Institute for Advanced Materials, Hubei Normal University, Huangshi 435002, China
}
\author{F. Steglich}
\affiliation{Center for Correlated Matter and Department of Physics,
Zhejiang University, Hangzhou 310058, China}
\affiliation{Max Planck Institute for Chemical Physics of Solids, Dresden, Germany}
\author{M. Smidman}
\email{msmidman@zju.edu.cn}
\affiliation{Center for Correlated Matter and Department of Physics,
Zhejiang University, Hangzhou 310058, China}
\affiliation{Zhejiang Province Key Laboratory of Quantum Technology and Device, Department of Physics, Zhejiang University, Hangzhou 310058, China}
\author{H. Q. Yuan}
\email{hqyuan@zju.edu.cn}
\affiliation{Center for Correlated Matter and Department of Physics,
Zhejiang University, Hangzhou 310058, China}
\affiliation{Zhejiang Province Key Laboratory of Quantum Technology and Device, Department of Physics, Zhejiang University, Hangzhou 310058, China}
\affiliation{State Key Laboratory of Silicon Materials, Zhejiang University, Hangzhou 310058, China}
\date{\today}
\addcontentsline{toc}{chapter}{Abstract}

\begin{abstract}
We report a study of isoelectronic chemical substitution in the recently discovered quantum critical ferromagnet CeRh$_6$Ge$_4$. Upon silicon-doping, the ferromagnetic ordering temperature of CeRh$_6$(Ge$_{1-x}$Si$_x$)$_4$ is continuously suppressed, and no transition is observed beyond $x_c$$\approx$0.125. Non-Fermi liquid behavior with $C/T \propto$log($T^*/T$) is observed close to $x_c$, indicating the existence of strong quantum fluctuations, while the $T$-linear behavior observed upon pressurizing the parent compound is absent in the resistivity, which appears to be a consequence of the disorder induced by silicon doping. Our findings provide evidence for the role played by disorder on  the unusual ferromagnetic quantum criticality in CeRh$_6$Ge$_4$, and provides further evidence for understanding the origin of this behavior.

\begin{description}
\item[PACS number(s)]
 74.70.-b, 71.45.Lr, 74.10.+v
\end{description}
\end{abstract}

\maketitle

\section{Introduction}

Quantum critical points (QCPs) are of extensive interest for understanding the physics of correlated electron systems, owing to the emergence of exotic quantum phenomena such as unconventional superconductivity and non-Fermi liquid behavior \cite{Pfleiderer_2009, Stewart_2001}. Heavy fermion metals are ideal systems for examining quantum criticality, since non-thermal parameters such as pressure, chemical doping or magnetic fields can effectively tune the relative strengths of the Ruderman-Kittel-Kasuya-Yosida interaction and the Kondo interaction, leading to the continuous suppression of a second-order phase transition to a QCP \cite{Weng_2016, Ruderman_1954, Kasuya_1956, Yosida_1957,Doniach_1977}.

QCPs have been extensively observed in antiferromagnetic systems \cite{Mathur_1998, Oeschler_2003, Custers_2003, Pietrus_1994}, but are rarely found in ferromagnetic materials, where the FM transition usually vanishes suddenly or converts to antiferromagnetic order with increasing the tuning parameter \cite{Brando_2016}. Recently, the stoichiometric heavy fermion compound CeRh$_6$Ge$_4$ was found to exhibit a FM QCP upon tuning with hydrostatic pressure \cite{Bin_2020, Kotegawa_2019}. Its ferromagnetic transition is continuously suppressed to zero at $p_c\approx$0.8~GPa, where there is also strange metal behavior with a linear temperature dependence of the resistivity and a logarithmic divergence of the specific heat coefficient \cite{Bin_2020}. These findings are in contrast to the anticipated absence of FM-QCPs in clean itinerant ferromagnetic systems \cite{Belitz_1999, Chubukov_2004}, and a local QCP scenario with localized ferromagnetism and magnetic anisotropy has been proposed to resolve this contradiction \cite{Bin_2020}. A localized nature of the 4$f$ electrons in CeRh$_6$Ge$_4$ was revealed in the subsequent study of quantum oscillations and density functional theory calculations \cite{Wang_2021}. Meanwhile, angle resolved photoemission spectroscopy \cite{Yi_2021}, ultrafast optical spectroscopy \cite{Pei_2021} and inelastic neutron scattering \cite{Shu_2021} provide evidence for anisotropic hybridization between the conduction and 4$f$ electrons in CeRh$_6$Ge$_4$.

\begin{figure*}[htbp]
\begin{center}
\includegraphics[width=15cm]{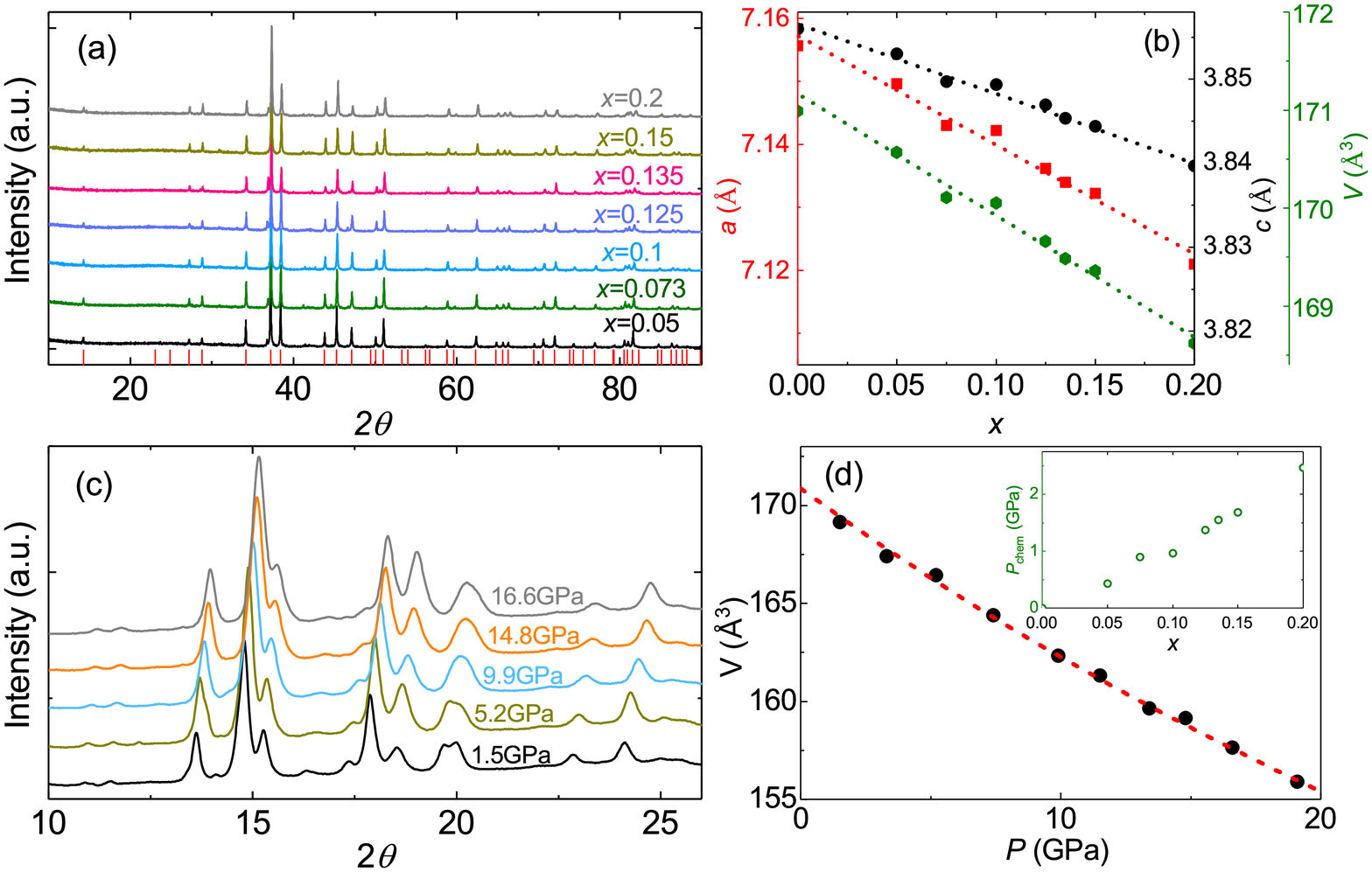}
\end{center}
\caption{(Color online) (a) X-ray diffraction (XRD) patterns of powdered CeRh$_6$(Ge$_{1-x}$Si$_x$)$_4$ polycrystalline samples, where the positions of the Bragg peaks corresponding to CeRh$_6$Ge$_4$ are marked by red vertical lines. (b) The lattice constants $a$, $c$ and the unit cell volume $V$ as a function of $x$ for CeRh$_6$(Ge$_{1-x}$Si$_x$)$_4$. Data for $a$, $c$ and $V$ are represented by red squares, black circles and olive hexagons, respectively, and the dashed lines are guides to the eye. (c) High pressure XRD patterns of CeRh$_6$Ge$_4$, with a monochromatic beam of wavelength 0.6199~\AA. (d) Unit cell volume of CeRh$_6$Ge$_4$ as a function of pressure, obtained from refining the high pressure powder XRD results. The dashed line represents the results from fitting with Eq.1. The inset shows the $x$ dependence of the chemical pressure $P_{chem}$.
}
\label{fig1}
\end{figure*}

The application of hydrostatic pressure to stoichiometric CeRh$_6$Ge$_4$ was vital for revealing the existence of a FM QCP without the introduction of disorder which is inherently present upon chemical doping. However, some experimental methods including thermal expansion and inelastic neutron scattering are extremely limited under pressure, which are essential for characterizing the QCP, such as the analysis of the divergent behavior of the Gr\"{u}neisen ratio \cite{Zhu_2003, Oeschler_2003, Steppke_2013} and searching for $E/T$ scaling in the dynamic susceptibility \cite{CeCu6_2000} at the QCP. Achieving a FM QCP via chemical substitution would expand the range of techniques which could be applied to probe the ferromagnetic quantum criticality in CeRh$_6$Ge$_4$. Moreover, the investigation of chemical substitution in CeRh$_6$Ge$_4$ allows for the investigation of the influence of disorder on the quantum critical behaviors in the vicinity of a FM QCP.

Here, in order to scrutinize these problems, we successfully synthesized a series of silicon-substituted CeRh$_6$Ge$_4$ samples. Their structural and physical properties are investigated using x-ray diffraction, electrical resistivity and specific-heat measurements. The reduction of the unit cell volume with increasing silicon-concentration demonstrates that chemical pressure has been applied relative to CeRh$_6$Ge$_4$. The $T-x$ phase diagram is constructed for CeRh$_6$(Ge$_{1-x}$Si$_x$)$_4$, where $T_C$ is suppressed upon silicon-doping, reaching at $x_c$$\approx$0.125. The logarithmic divergence of $C/T$ observed close to $x_c$ indicates the existence of strong quantum fluctuations in silicon-doped CeRh$_6$Ge$_4$, but the $T$-linear temperature dependence of the resistivity observed for the stoichiometric compound \cite{Bin_2020} is found to be absent in the resistivity, likely due to the influence of disorder.

\section{Experimental details}

Polycrystalline samples of CeRh$_6$(Ge$_{1-x}$Si$_x$)$_4$ were synthesized by arc-melting in a titanium-gettered argon atmosphere, with a stoichiometric composition of Ce~ingot ($99.9\%$), Rh~ingot ($99.95\%$) and Ge$_{1-x}$Si$_x$ ingot \cite{Matsuoka_2015}. The precursors Ge$_{1-x}$Si$_x$ ingots were prepared by alloying Ge ingot ($99.999\%$) with Si ingot ($99.999\%$) in the nominal atomic ratio. The obtained CeRh$_6$(Ge$_{1-x}$Si$_x$)$_4$ ingots were wrapped in Ta foil and sealed in evacuated quartz tubes before being annealed at 1100$~^\circ$C. The composition of all doped samples were determined by energy dispersive x-ray spectroscopy. The crystal structure was characterized using powder x-ray diffraction, and the high pressure x-ray diffraction measurements of stoichiometric CeRh$_6$Ge$_4$ were performed at the 4W2 beamline at the Beijing Synchrotron Radiation Facility (BSRF) with a monochromatic beam of wavelength 0.6199~\AA. The heat capacity was measured down to 0.4~K using a Quantum Design Physical Property Measurement System (PPMS) with a $^3$He insert. The resistivity measurements were measured down to 1.8~K and 0.4~K in a PPMS and $^3$He refrigerator, respectively.

\section{results}

\begin{figure}[h]
\includegraphics[width=7.5cm]{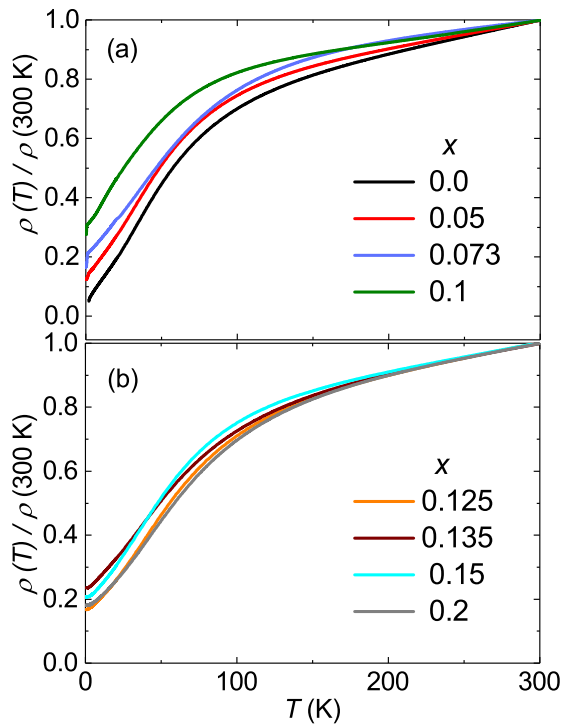}
\caption{(Color online) Temperature dependence of the normalized resistivity $\rho(T)/\rho(300K)$ for samples with doping concentrations (a) $x\leq$0.1 and (b) $x\geq$0.125.
}
\label{fig2}
\end{figure}

Powder XRD patterns for CeRh$_6$(Ge$_{1-x}$Si$_x$)$_4$ are shown in Fig.~1(a), where the peak positions can be well indexed on the basis of the hexagonal CeRh$_6$Ge$_4$ structure \cite{Matsuoka_2015}. In order to reveal the evolution of the lattice with silicon substitution, the refined lattice constants $a(x)$, $c(x)$ and unit cell volume $V(x)$ are displayed in Fig.~1(b). Up to $x$=0.2, all the lattice constants $a$, $c$ and unit cell volume $V$ decrease linearly as a function of $x$, which is due to the isoelectronic substitution of the larger germanium atom with the smaller silicon. The $a/c$ ratio is almost independent of $x$, with a $a/c$=1.85, which is similar to that reported for CeRh$_6$Ge$_4$ single crystals \cite{Daniel_2012}. The continuous lattice contraction and constant value of $a/c$ ratio indicates the absence of a change in crystal structure of the CeRh$_6$(Ge$_{1-x}$Si$_x$)$_4$.

In order to reveal the pressure evolution of the lattice constants in CeRh$_6$Ge$_4$ and to estimate the equivalent chemical pressure induced by Si/Ge substitution, high-pressure XRD measurements of CeRh$_6$Ge$_4$ were conducted in the pressure range 1.5 to 16.6 GPa, and the high-pressure XRD patterns are displayed in Fig.~1(c). The obtained volume $V$ as a function of pressure is displayed in Fig.~1(d), where the volume continuously decreases upon applying pressure, indicating the absence of a pressure-induced structural transition in CeRh$_6$Ge$_4$. The data are well fitted by \cite{Birch_1947}
\begin{equation}
P_{chem}=\frac{3B_0}{2}[(\frac{V_0}{V})^{\frac{7}{3}}-(\frac{V_0}{V})^{\frac{5}{3}}],
\end{equation}
giving rise to a bulk modulus $B_0$=174(3) and zero pressure volume $V_0$=170.9(2)~\AA$^3$. The isoelectronic substitution of silicon for germanium has a similar effect to hydrostatic pressure, both resulting in a contraction of the lattice. In order to reveal the chemical pressure of silicon-doping, the $P_{chem}$ as function of $x$ is shown in the inset, where $P_{chem}$ is estimated from comparing the values of the lattice volume under silicon-doping and pressure.

\begin{figure}[h]
\includegraphics[width=7.5cm]{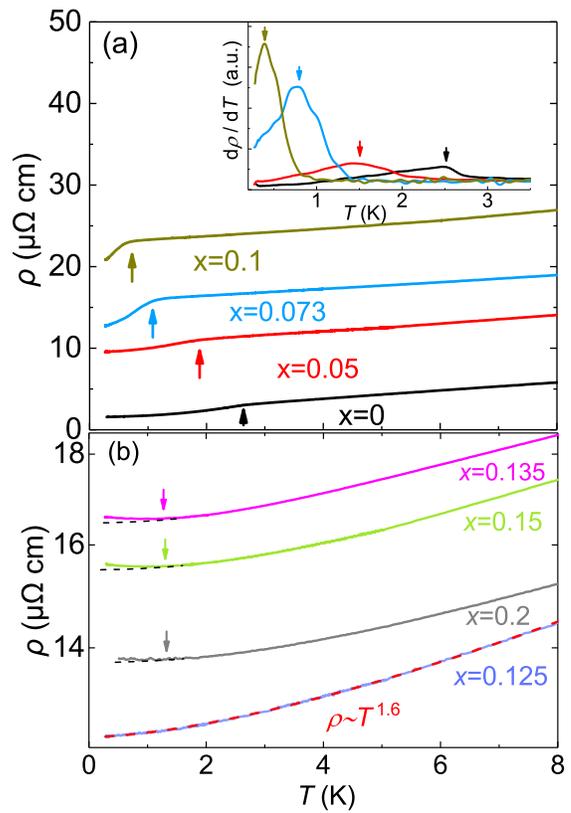}
\caption{(Color online) (a) The low-temperature resistivity $\rho$($T$) for $x\leq$0.1. The inset shows the derivative d$\rho$/d$T$ for $x\leq$0.1. The arrows correspond to the ferromagnetic transition at $T_C$. (b) Low temperature resistivity $\rho$($T$) for $x\geq$0.125. The red dashed lines show fits to non-Fermi liquid behavior with $\rho$($T$)$\sim$$T^{1.6}$, the arrows mark the upturn behavior at low temperature, and the black dashed lines correspond to a Fermi liquid behavior  from which the data deviates.
}
\label{fig3}
\end{figure}

\begin{figure}[h]
\includegraphics[width=8.5cm]{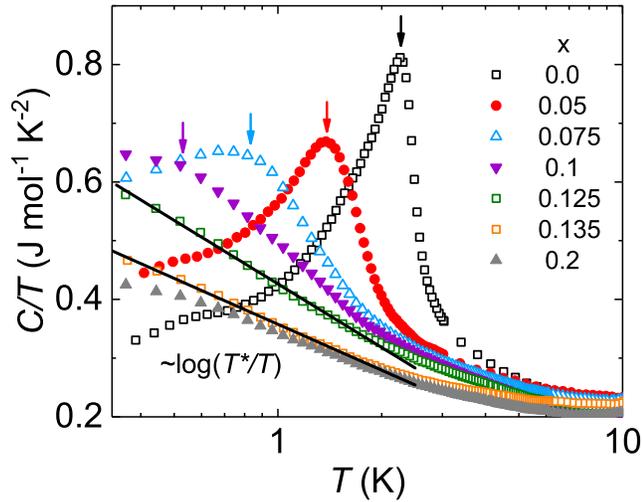}
\caption{(Color online) Temperature dependence of the specific heat as $C/T$ for $x\leq$0.2. The arrows mark the position of the Curie temperature $T_C$. The black solid lines correspond to the log($T^*/T$) divergence of $C/T$.
}
\label{fig4}
\end{figure}

The temperature dependence of the normalized resistivity $\rho(T)/\rho$(300K) for CeRh$_6$(Ge$_{1-x}$Si$_x$)$_4$ is shown in Figs.~2(a) and 2(b). It is clear that the residual resistivity ratio $RRR$ decreases with doping for $x\leq$0.1, as seen in Fig.~2(a), and the increase of residual resistivity $\rho_0$ with $x$ is much greater than that of stoichiometric CeRh$_6$Ge$_4$ under pressure \cite{Bin_2020}, which corresponds to the disorder effect induced by silicon-substitution.

Figure~3 displays the resistivity $\rho$($T$) and corresponding derivatives d$\rho$/d$T$ at temperatures below 8~K for various $x$. As seen in Fig.~3(a), both $\rho$($T$) and d$\rho$/d$T$ for $x$=0 show an anomaly at $T_C$, which corresponds to the ferromagnetic ordering transition reported previously \cite{Matsuoka_2015,Bin_2020}. With increasing $x$, the transition is suppressed to lower temperatures, and disappears beyond $x$=0.1. For $x$=0.125, there is no transition observed, and non-Fermi liquid is observed in $\rho$($T$) down to the lowest measured temperature. The non Fermi liquid behavior were analyzed using the expression $\rho$$\sim$$T^{1.6}$, as shown in Fig.~3(b), which is different from the $T$-linear behavior observed at the FM QCP in stoichiometric CeRh$_6$Ge$_4$ \cite{Bin_2020}. For $x\geq$0.135, with decreasing temperature below 1.0~K, there is a weak upturn of $\rho$($T$), as marked by the arrows in Fig.~3(b). These upturn behaviors may be a consequence of the enhanced disorder effect induced by silicon-doping.

The low-temperature $C/T$ is displayed in Fig.~4. For $x$$\leq$0.1, a sharp transition is observed in $C/T$, corresponding to the ferromagnetic transition, where $T_C$ gradually shifts to lower temperature with increasing $x$, and no transition is detected for $x$$\geq$0.125, being consistent with measurements of $\rho$($T$) in Fig.~3(a). A logarithmic temperature dependence of $C/T$ is also observed for $x$=0.125 and 0.135, indicating the existence of strong quantum fluctuations for these concentrations, as observed at the FM QCP in stoichiometric CeRh$_6$Ge$_4$ \cite{Bin_2020}. In the paramagnetic state beyond $x$=0.135, the divergent behavior of $C/T$ is gradually suppressed upon further increasing $x$, which likely indicates the weakening of these fluctuations.

\begin{figure}[h]
\includegraphics[width=8cm]{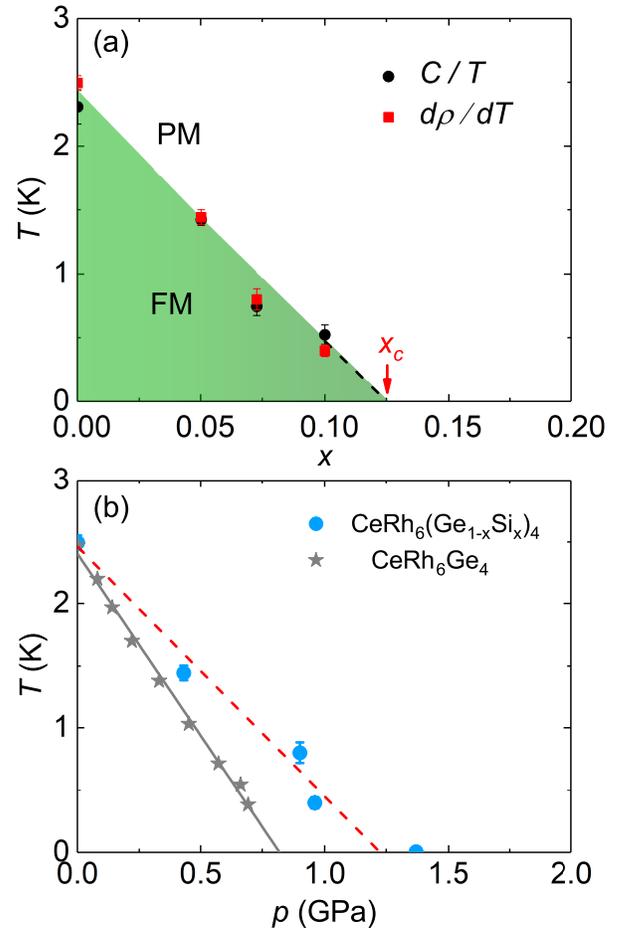}
\caption{(Color online)  (a) $T-x$ phase diagram of CeRh$_6$(Ge$_{1-x}$Si$_x$)$_4$. The black circles and red squares represent $T_C$ derived from the specific heat and resistivity, respectively, and the errors correspond to the scattering of the low-$T$ data. (b) $T-P_{chem}$ phase diagram for CeRh$_6$(Ge$_{1-x}$Si$_x$)$_4$ and $T-P$ phase diagram for pressurized CeRh$_6$Ge$_4$. The evolution of $T_C$ with pressure for stoichiometric CeRh$_6$Ge$_4$ is obtained from Ref. \cite{Bin_2020}.
}
\label{fig5}
\end{figure}

Based on the measurements of the specific heat and resistivity of CeRh$_6$(Ge$_{1-x}$Si$_x$)$_4$, the $T-x$ phase diagram is summarized in Fig.~5(a). The values of $T_C$ obtained from $C/T$ and $\rho(T)$ are all consistent, where $T_C$ is continuously suppressed with increasing $x_c$, before being no longer observed beyond $x$=0.125. The evolution of magnetism upon silicon-doping is similar to the $T-P$ phase diagram in stoichiometric CeRh$_6$Ge$_4$ \cite{Bin_2020}, where $T_C$ is linearly suppressed under pressure. As highlighted by the dashed line in Fig.~5(a), a critical concentration $x_c\approx$0.125 can be obtained from a linear extrapolation of the slope of $T_C(x)$.

\begin{figure}[h]
\includegraphics[width=8.5cm]{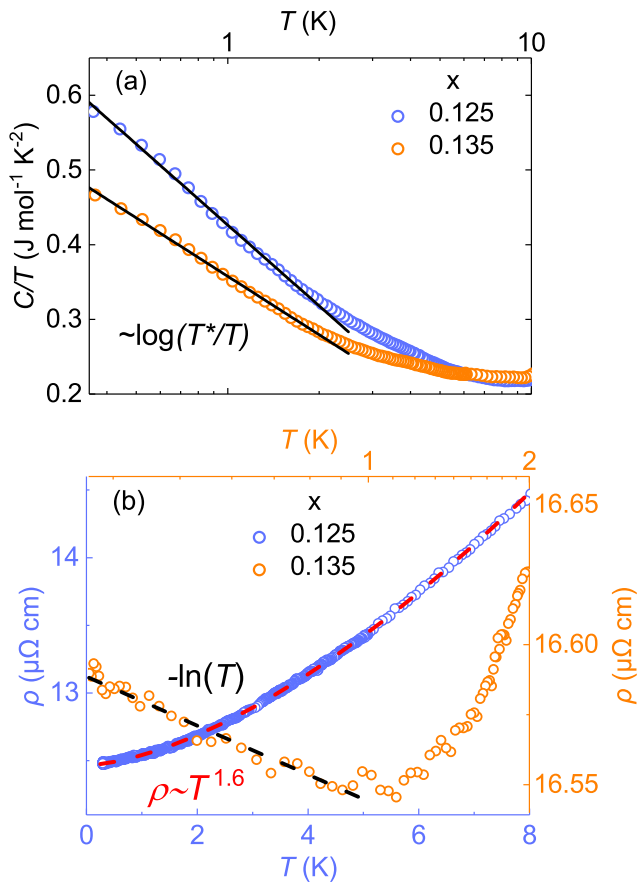}
\caption{(Color online)  (a) Temperature dependence of $C/T$ for $x$=0.125 and 0.135. The black solid lines correspond to the log($T^*/T$) divergence of $C/T$. (b) $\rho(T)$ for $x$=0.125 and $x$=0.135. The red dashed line corresponds to a fit with $\rho$$\propto$ $T^{1.6}$, and the black dashed line represents a fit to $\rho(T)$$\propto$-ln$T$.
}
\label{fig6}
\end{figure}

\section{DISCUSSION}

The $T_C$ of CeRh$_6$(Ge$_{1-x}$Si$_x$)$_4$ is continuously suppressed upon silicon-doping, until it is no longer observed at a critical concentration of $x_c\approx$0.125. Since the effects of isoelectronic silicon-doping is primarily a pressure effect, this suggests that the mechanism for the suppression of $T_C$ is similar to that observed for hydrostatic pressure, where a ferromagnetic quantum critical point and strange metal behavior is observed \cite{Bin_2020}. In order to compare the effects of chemical and hydrostatic pressures in CeRh$_6$Ge$_4$, the $T-P$ phase diagram of CeRh$_6$(Ge$_{1-x}$Si$_x$)$_4$ and CeRh$_6$Ge$_4$ under pressure is summarized in Fig.~5(b). The evolution of $T_C$ of chemical pressure $P_{chem}$ for CeRh$_6$(Ge$_{1-x}$Si$_x$)$_4$ was derived from the inset of Fig.~1(d) and Fig.5(a), and the evolution of $T_C$ with pressure for stoichiometric CeRh$_6$Ge$_4$ is from Ref. \cite{Bin_2020}. All the $T_C$ are continuously suppressed with increasing pressure $p$, indicating that the chemical pressure induced by silicon-doping indeed has a similar effect to hydrostatic pressure on CeRh$_6$Ge$_4$. The difference between the data derived from applying hydrostatic and chemical pressure, may be due to the pressure for the XRD measurements being measured at room temperature, while the pressure for transport measurements was determined at low temperatures.

As displayed in Fig.~6(a), a logarithmic temperature dependence of $C/T$ is observed for $x$=0.125 and 0.135, indicating the existence of strong quantum fluctuations in the vicinity of $x_c$. The concentration dependence of $T_C$ and logarithmic divergence of the $C/T$ provide evidence for the existence of a FM QCP at $x_c$$\approx$0.125. Such specific heat behavior with $C/T$$\propto$log$(T^*/T)$, has been observed at the pressure-induced FM-QCP of CeRh$_6$Ge$_4$ \cite{Bin_2020} as well as in La-doped CeRh$_6$Ge$_4$ \cite{Cheng_2021}. However, while $T$-linear behavior of $\rho(T)$ is observed in pressurized stoichiometric CeRh$_6$Ge$_4$ at the FM QCP, it is not found in CeRh$_6$(Ge$_{1-x}$Si$_x$)$_4$ upon the suppression of FM order. As seen in Fig.~6(b), a $T^{1.6}$ behavior dependence of $\rho(T)$ is observed at $x$=0.125, while at a slightly higher concentration of $x$=0.135, a low temperature upturn with $\rho$$\sim$ln$(1/T)$ behavior is observed. The non-Fermi liquid behavior with $T^{1.6}$ might arise from the interplay of disorder and spin fluctuations near a QCP as discussed for antiferromagnetic systems \cite{Rosch_1999}, where the resistivity varies as $T^\alpha$ with 1$\leq$$\alpha$$\leq$1.5, depending on the degree of disorder. The $\rho$$\sim$ln$(1/T)$ observed for high doping concentration samples in Si-doped CeRh$_6$Ge$_4$ is different from the $\rho$$\sim$-$AT$ reported in La-doped CeRh$_6$Ge$_4$ \cite{Cheng_2021}, which can be described in terms of a disordered Kondo model \cite{Stewart_2001}. However, a similar upturn is observed in CePd$_{1-x}$ Rh$_x$, which is attributed to moments not being fully screened by the Kondo effect \cite{Hilbert_2007}. Additional studies are necessary to determine the role played by disorder on the quantum critical behaviors and to directly confirm the presence of quantum criticality in CeRh$_6$(Ge$_{1-x}$Si$_x$)$_4$.

\section{Summary}

In conclusion, we report an investigation of silicon substitution on the quantum critical ferromagnet CeRh$_6$Ge$_4$. The suppression of the Curie temperature $T_C$ upon silicon-doping and logarithmic divergence of $C/T$ suggests the possible presence of a QCP at $x_c\approx$0.125. Here the reduction of $T_C$  with isoelectronic substitution of Si for Ge is attributed to the chemical pressure effect, where both hydrostatic and chemical pressures reduce the unit cell volume. At $x_c\approx$0.125, a log($T^*/T$) behavior is observed in $C/T$, which is similar to the case of stoichiometric CeRh$_6$Ge$_4$ under pressure \cite{Bin_2020}, but the resistivity deviates from the $T$-linear behavior observed for the stoichiometric compound, indicating the influence of disorder. In future, thermal expansion and neutron scattering are necessary to characterize the nature of the potential quantum criticality in CeRh$_6$(Ge$_{1-x}$Si$_x$)$_4$.

\section{acknowledgments}

This work was supported by the National Natural Science Foundation of China (No.~12034017, No.~11974306 and No.~12174332), the National Key R\&D Program of China (No.~2017YFA0303100), the Key R\&D Program of Zhejiang Province, China (2021C01002) and the Hubei Provincial Natural Science Foundation of China with Grant No.~2021CFB220.


\begin{thebibliography}{29}%
\makeatletter
\providecommand \@ifxundefined [1]{%
 \@ifx{#1\undefined}
}%
\providecommand \@ifnum [1]{%
 \ifnum #1\expandafter \@firstoftwo
 \else \expandafter \@secondoftwo
 \fi
}%
\providecommand \@ifx [1]{%
 \ifx #1\expandafter \@firstoftwo
 \else \expandafter \@secondoftwo
 \fi
}%
\providecommand \natexlab [1]{#1}%
\providecommand \enquote  [1]{``#1''}%
\providecommand \bibnamefont  [1]{#1}%
\providecommand \bibfnamefont [1]{#1}%
\providecommand \citenamefont [1]{#1}%
\providecommand \href@noop [0]{\@secondoftwo}%
\providecommand \href [0]{\begingroup \@sanitize@url \@href}%
\providecommand \@href[1]{\@@startlink{#1}\@@href}%
\providecommand \@@href[1]{\endgroup#1\@@endlink}%
\providecommand \@sanitize@url [0]{\catcode `\\12\catcode `\$12\catcode
  `\&12\catcode `\#12\catcode `\^12\catcode `\_12\catcode `\%12\relax}%
\providecommand \@@startlink[1]{}%
\providecommand \@@endlink[0]{}%
\providecommand \url  [0]{\begingroup\@sanitize@url \@url }%
\providecommand \@url [1]{\endgroup\@href {#1}{\urlprefix }}%
\providecommand \urlprefix  [0]{URL }%
\providecommand \Eprint [0]{\href }%
\providecommand \doibase [0]{http://dx.doi.org/}%
\providecommand \selectlanguage [0]{\@gobble}%
\providecommand \bibinfo  [0]{\@secondoftwo}%
\providecommand \bibfield  [0]{\@secondoftwo}%
\providecommand \translation [1]{[#1]}%
\providecommand \BibitemOpen [0]{}%
\providecommand \bibitemStop [0]{}%
\providecommand \bibitemNoStop [0]{.\EOS\space}%
\providecommand \EOS [0]{\spacefactor3000\relax}%
\providecommand \BibitemShut  [1]{\csname bibitem#1\endcsname}%
\let\auto@bib@innerbib\@empty
\bibitem [{\citenamefont {Pfleiderer}(2009)}]{Pfleiderer_2009}%
  \BibitemOpen
  \bibfield  {author} {\bibinfo {author} {\bibfnamefont {C.}~\bibnamefont
  {Pfleiderer}},\ }\bibfield  {title} {\enquote {\bibinfo {title}
  {Superconducting phases of $f$-electron compounds},}\ }\href {\doibase
  10.1103/RevModPhys.81.1551} {\bibfield  {journal} {\bibinfo  {journal} {Rev.
  Mod. Phys.}\ }\textbf {\bibinfo {volume} {81}},\ \bibinfo {pages}
  {1551--1624} (\bibinfo {year} {2009})}\BibitemShut {NoStop}%
\bibitem [{\citenamefont {Stewart}(2001)}]{Stewart_2001}%
  \BibitemOpen
  \bibfield  {author} {\bibinfo {author} {\bibfnamefont {G.~R.}\ \bibnamefont
  {Stewart}},\ }\bibfield  {title} {\enquote {\bibinfo {title}
  {Non-fermi-liquid behavior in d- and f -electron metals},}\ }\href {\doibase
  10.1103/RevModPhys.73.797} {\bibfield  {journal} {\bibinfo  {journal} {Rev.
  Mod. Phys.}\ }\textbf {\bibinfo {volume} {73}},\ \bibinfo {pages} {797--855}
  (\bibinfo {year} {2001})}\BibitemShut {NoStop}%
\bibitem [{\citenamefont {Weng}\ \emph {et~al.}(2016)\citenamefont {Weng},
  \citenamefont {Smidman}, \citenamefont {Jiao}, \citenamefont {Lu},\ and\
  \citenamefont {Yuan}}]{Weng_2016}%
  \BibitemOpen
  \bibfield  {author} {\bibinfo {author} {\bibfnamefont {Z.~F.}\ \bibnamefont
  {Weng}}, \bibinfo {author} {\bibfnamefont {M.}~\bibnamefont {Smidman}},
  \bibinfo {author} {\bibfnamefont {L.}~\bibnamefont {Jiao}}, \bibinfo {author}
  {\bibfnamefont {X.}~\bibnamefont {Lu}}, \ and\ \bibinfo {author}
  {\bibfnamefont {H.~Q.}\ \bibnamefont {Yuan}},\ }\bibfield  {title} {\enquote
  {\bibinfo {title} {Multiple quantum phase transitions and superconductivity
  in {Ce}-based heavy fermions},}\ }\href {\doibase
  10.1088/0034-4885/79/9/094503} {\bibfield  {journal} {\bibinfo  {journal}
  {Reports on Progress in Physics}\ }\textbf {\bibinfo {volume} {79}},\
  \bibinfo {pages} {094503} (\bibinfo {year} {2016})}\BibitemShut {NoStop}%
\bibitem [{\citenamefont {Ruderman}\ and\ \citenamefont
  {Kittel}(1954)}]{Ruderman_1954}%
  \BibitemOpen
  \bibfield  {author} {\bibinfo {author} {\bibfnamefont {M.~A.}\ \bibnamefont
  {Ruderman}}\ and\ \bibinfo {author} {\bibfnamefont {C.}~\bibnamefont
  {Kittel}},\ }\bibfield  {title} {\enquote {\bibinfo {title} {Indirect
  exchange coupling of nuclear magnetic moments by conduction electrons},}\
  }\href {\doibase 10.1103/PhysRev.96.99} {\bibfield  {journal} {\bibinfo
  {journal} {Phys. Rev.}\ }\textbf {\bibinfo {volume} {96}},\ \bibinfo {pages}
  {99--102} (\bibinfo {year} {1954})}\BibitemShut {NoStop}%
\bibitem [{\citenamefont {Kasuya}(1956)}]{Kasuya_1956}%
  \BibitemOpen
  \bibfield  {author} {\bibinfo {author} {\bibfnamefont {T.}~\bibnamefont
  {Kasuya}},\ }\bibfield  {title} {\enquote {\bibinfo {title} {A theory of
  metallic ferro- and antiferromagnetism on {Zener's} model},}\ }\href
  {\doibase 10.1143/PTP.16.45} {\bibfield  {journal} {\bibinfo  {journal}
  {Progress of Theoretical Physics}\ }\textbf {\bibinfo {volume} {16}},\
  \bibinfo {pages} {45--57} (\bibinfo {year} {1956})}\BibitemShut {NoStop}%
\bibitem [{\citenamefont {Yosida}(1957)}]{Yosida_1957}%
  \BibitemOpen
  \bibfield  {author} {\bibinfo {author} {\bibfnamefont {K.}~\bibnamefont
  {Yosida}},\ }\bibfield  {title} {\enquote {\bibinfo {title} {Magnetic
  properties of {Cu-Mn} alloys},}\ }\href {\doibase 10.1103/PhysRev.106.893}
  {\bibfield  {journal} {\bibinfo  {journal} {Phys. Rev.}\ }\textbf {\bibinfo
  {volume} {106}},\ \bibinfo {pages} {893--898} (\bibinfo {year}
  {1957})}\BibitemShut {NoStop}%
\bibitem [{\citenamefont {Doniach}(1977)}]{Doniach_1977}%
  \BibitemOpen
  \bibfield  {author} {\bibinfo {author} {\bibfnamefont {S.}~\bibnamefont
  {Doniach}},\ }\bibfield  {title} {\enquote {\bibinfo {title} {The {K}ondo
  lattice and weak antiferromagnetism},}\ }\href {\doibase
  10.1016/0378-4363(77)90190-5} {\bibfield  {journal} {\bibinfo  {journal}
  {Physica B+C}\ }\textbf {\bibinfo {volume} {91}},\ \bibinfo {pages} {231 --
  234} (\bibinfo {year} {1977})}\BibitemShut {NoStop}%
\bibitem [{\citenamefont {Mathur}\ \emph {et~al.}(1998)\citenamefont {Mathur},
  \citenamefont {Grosche}, \citenamefont {Julian}, \citenamefont {Walker},
  \citenamefont {Freye}, \citenamefont {Haselwimmer},\ and\ \citenamefont
  {Lonzarich}}]{Mathur_1998}%
  \BibitemOpen
  \bibfield  {author} {\bibinfo {author} {\bibfnamefont {N.~D.}\ \bibnamefont
  {Mathur}}, \bibinfo {author} {\bibfnamefont {F.~M.}\ \bibnamefont {Grosche}},
  \bibinfo {author} {\bibfnamefont {S.~R.}\ \bibnamefont {Julian}}, \bibinfo
  {author} {\bibfnamefont {I.~R.}\ \bibnamefont {Walker}}, \bibinfo {author}
  {\bibfnamefont {D.~M.}\ \bibnamefont {Freye}}, \bibinfo {author}
  {\bibfnamefont {R.~K.~W.}\ \bibnamefont {Haselwimmer}}, \ and\ \bibinfo
  {author} {\bibfnamefont {G.~G.}\ \bibnamefont {Lonzarich}},\ }\bibfield
  {title} {\enquote {\bibinfo {title} {Magnetically mediated superconductivity
  in heavy fermion compound},}\ }\href {\doibase 10.1038/27838} {\bibfield
  {journal} {\bibinfo  {journal} {Nature}\ }\textbf {\bibinfo {volume} {394}},\
  \bibinfo {pages} {39} (\bibinfo {year} {1998})}\BibitemShut {NoStop}%
\bibitem [{\citenamefont {K\"uchler}\ \emph {et~al.}(2003)\citenamefont
  {K\"uchler}, \citenamefont {Oeschler}, \citenamefont {Gegenwart},
  \citenamefont {Cichorek}, \citenamefont {Neumaier}, \citenamefont {Tegus},
  \citenamefont {Geibel}, \citenamefont {Mydosh}, \citenamefont {Steglich},
  \citenamefont {Zhu},\ and\ \citenamefont {Si}}]{Oeschler_2003}%
  \BibitemOpen
  \bibfield  {author} {\bibinfo {author} {\bibfnamefont {R.}~\bibnamefont
  {K\"uchler}}, \bibinfo {author} {\bibfnamefont {N.}~\bibnamefont {Oeschler}},
  \bibinfo {author} {\bibfnamefont {P.}~\bibnamefont {Gegenwart}}, \bibinfo
  {author} {\bibfnamefont {T.}~\bibnamefont {Cichorek}}, \bibinfo {author}
  {\bibfnamefont {K.}~\bibnamefont {Neumaier}}, \bibinfo {author}
  {\bibfnamefont {O.}~\bibnamefont {Tegus}}, \bibinfo {author} {\bibfnamefont
  {C.}~\bibnamefont {Geibel}}, \bibinfo {author} {\bibfnamefont {J.~A.}\
  \bibnamefont {Mydosh}}, \bibinfo {author} {\bibfnamefont {F.}~\bibnamefont
  {Steglich}}, \bibinfo {author} {\bibfnamefont {L.}~\bibnamefont {Zhu}}, \
  and\ \bibinfo {author} {\bibfnamefont {Q.}~\bibnamefont {Si}},\ }\bibfield
  {title} {\enquote {\bibinfo {title} {Divergence of the gr\"uneisen ratio at
  quantum critical points in heavy fermion metals},}\ }\href {\doibase
  10.1103/PhysRevLett.91.066405} {\bibfield  {journal} {\bibinfo  {journal}
  {Phys. Rev. Lett.}\ }\textbf {\bibinfo {volume} {91}},\ \bibinfo {pages}
  {066405} (\bibinfo {year} {2003})}\BibitemShut {NoStop}%
\bibitem [{\citenamefont {Custers}\ \emph {et~al.}(2003)\citenamefont
  {Custers}, \citenamefont {Gegenwart}, \citenamefont {Wilhelm}, \citenamefont
  {Neumaier}, \citenamefont {Tokiwa}, \citenamefont {Trovarelli}, \citenamefont
  {Geibel}, \citenamefont {Steglich}, \citenamefont {Pepin},\ and\
  \citenamefont {Coleman}}]{Custers_2003}%
  \BibitemOpen
  \bibfield  {author} {\bibinfo {author} {\bibfnamefont {J.}~\bibnamefont
  {Custers}}, \bibinfo {author} {\bibfnamefont {P.}~\bibnamefont {Gegenwart}},
  \bibinfo {author} {\bibfnamefont {H.}~\bibnamefont {Wilhelm}}, \bibinfo
  {author} {\bibfnamefont {K.}~\bibnamefont {Neumaier}}, \bibinfo {author}
  {\bibfnamefont {Y.}~\bibnamefont {Tokiwa}}, \bibinfo {author} {\bibfnamefont
  {O.}~\bibnamefont {Trovarelli}}, \bibinfo {author} {\bibfnamefont
  {C.}~\bibnamefont {Geibel}}, \bibinfo {author} {\bibfnamefont
  {F.}~\bibnamefont {Steglich}}, \bibinfo {author} {\bibfnamefont
  {C.}~\bibnamefont {Pepin}}, \ and\ \bibinfo {author} {\bibfnamefont
  {P.}~\bibnamefont {Coleman}},\ }\bibfield  {title} {\enquote {\bibinfo
  {title} {The break-up of heavy electrons at a quantum critical point},}\
  }\href {\doibase 10.1038/nature01774} {\bibfield  {journal} {\bibinfo
  {journal} {Nature}\ }\textbf {\bibinfo {volume} {424}},\ \bibinfo {pages}
  {524} (\bibinfo {year} {2003})}\BibitemShut {NoStop}%
\bibitem [{\citenamefont {L\"ohneysen}\ \emph {et~al.}(1994)\citenamefont
  {L\"ohneysen}, \citenamefont {Pietrus}, \citenamefont {Portisch},
  \citenamefont {Schlager}, \citenamefont {Schr\"oder}, \citenamefont {Sieck},\
  and\ \citenamefont {Trappmann}}]{Pietrus_1994}%
  \BibitemOpen
  \bibfield  {author} {\bibinfo {author} {\bibfnamefont {H.~v.}\ \bibnamefont
  {L\"ohneysen}}, \bibinfo {author} {\bibfnamefont {T.}~\bibnamefont
  {Pietrus}}, \bibinfo {author} {\bibfnamefont {G.}~\bibnamefont {Portisch}},
  \bibinfo {author} {\bibfnamefont {H.~G.}\ \bibnamefont {Schlager}}, \bibinfo
  {author} {\bibfnamefont {A.}~\bibnamefont {Schr\"oder}}, \bibinfo {author}
  {\bibfnamefont {M.}~\bibnamefont {Sieck}}, \ and\ \bibinfo {author}
  {\bibfnamefont {T.}~\bibnamefont {Trappmann}},\ }\bibfield  {title} {\enquote
  {\bibinfo {title} {Non-fermi-liquid behavior in a heavy-fermion alloy at a
  magnetic instability},}\ }\href {\doibase 10.1103/PhysRevLett.72.3262}
  {\bibfield  {journal} {\bibinfo  {journal} {Phys. Rev. Lett.}\ }\textbf
  {\bibinfo {volume} {72}},\ \bibinfo {pages} {3262--3265} (\bibinfo {year}
  {1994})}\BibitemShut {NoStop}%
\bibitem [{\citenamefont {Brando}\ \emph {et~al.}(2016)\citenamefont {Brando},
  \citenamefont {Belitz}, \citenamefont {Grosche},\ and\ \citenamefont
  {Kirkpatrick}}]{Brando_2016}%
  \BibitemOpen
  \bibfield  {author} {\bibinfo {author} {\bibfnamefont {M.}~\bibnamefont
  {Brando}}, \bibinfo {author} {\bibfnamefont {D.}~\bibnamefont {Belitz}},
  \bibinfo {author} {\bibfnamefont {F.~M.}\ \bibnamefont {Grosche}}, \ and\
  \bibinfo {author} {\bibfnamefont {T.~R.}\ \bibnamefont {Kirkpatrick}},\
  }\bibfield  {title} {\enquote {\bibinfo {title} {Metallic quantum
  ferromagnets},}\ }\href {\doibase 10.1103/RevModPhys.88.025006} {\bibfield
  {journal} {\bibinfo  {journal} {Rev. Mod. Phys.}\ }\textbf {\bibinfo {volume}
  {88}},\ \bibinfo {pages} {025006} (\bibinfo {year} {2016})}\BibitemShut
  {NoStop}%
\bibitem [{\citenamefont {Shen}\ \emph {et~al.}(2020)\citenamefont {Shen},
  \citenamefont {Zhang}, \citenamefont {Komijani}, \citenamefont {Nicklas},
  \citenamefont {Borth}, \citenamefont {Wang}, \citenamefont {Chen},
  \citenamefont {Nie}, \citenamefont {Li}, \citenamefont {Lu}, \citenamefont
  {Lee}, \citenamefont {Smidman}, \citenamefont {Steglich}, \citenamefont
  {Coleman},\ and\ \citenamefont {Yuan}}]{Bin_2020}%
  \BibitemOpen
  \bibfield  {author} {\bibinfo {author} {\bibfnamefont {Bin}\ \bibnamefont
  {Shen}}, \bibinfo {author} {\bibfnamefont {Yongjun}\ \bibnamefont {Zhang}},
  \bibinfo {author} {\bibfnamefont {Yashar}\ \bibnamefont {Komijani}}, \bibinfo
  {author} {\bibfnamefont {Michael}\ \bibnamefont {Nicklas}}, \bibinfo {author}
  {\bibfnamefont {Robert}\ \bibnamefont {Borth}}, \bibinfo {author}
  {\bibfnamefont {An}~\bibnamefont {Wang}}, \bibinfo {author} {\bibfnamefont
  {Ye}~\bibnamefont {Chen}}, \bibinfo {author} {\bibfnamefont {Zhiyong}\
  \bibnamefont {Nie}}, \bibinfo {author} {\bibfnamefont {Rui}\ \bibnamefont
  {Li}}, \bibinfo {author} {\bibfnamefont {Xin}\ \bibnamefont {Lu}}, \bibinfo
  {author} {\bibfnamefont {Hanoh}\ \bibnamefont {Lee}}, \bibinfo {author}
  {\bibfnamefont {Michael}\ \bibnamefont {Smidman}}, \bibinfo {author}
  {\bibfnamefont {Frank}\ \bibnamefont {Steglich}}, \bibinfo {author}
  {\bibfnamefont {Piers}\ \bibnamefont {Coleman}}, \ and\ \bibinfo {author}
  {\bibfnamefont {Huiqiu}\ \bibnamefont {Yuan}},\ }\bibfield  {title} {\enquote
  {\bibinfo {title} {Strange-metal behaviour in a pure ferromagnetic kondo
  lattice},}\ }\href {\doibase 10.1038/s41586-020-2052-z} {\bibfield  {journal}
  {\bibinfo  {journal} {Nature}\ }\textbf {\bibinfo {volume} {579}},\ \bibinfo
  {pages} {51} (\bibinfo {year} {2020})}\BibitemShut {NoStop}%
\bibitem [{\citenamefont {Kotegawa}\ \emph {et~al.}(2019)\citenamefont
  {Kotegawa}, \citenamefont {Matsuoka}, \citenamefont {Uga}, \citenamefont
  {Takemura}, \citenamefont {Manago}, \citenamefont {Chikuchi}, \citenamefont
  {Sugawara}, \citenamefont {Tou},\ and\ \citenamefont
  {Harima}}]{Kotegawa_2019}%
  \BibitemOpen
  \bibfield  {author} {\bibinfo {author} {\bibfnamefont {Hisashi}\ \bibnamefont
  {Kotegawa}}, \bibinfo {author} {\bibfnamefont {Eiichi}\ \bibnamefont
  {Matsuoka}}, \bibinfo {author} {\bibfnamefont {Toshiaki}\ \bibnamefont
  {Uga}}, \bibinfo {author} {\bibfnamefont {Masaki}\ \bibnamefont {Takemura}},
  \bibinfo {author} {\bibfnamefont {Masahiro}\ \bibnamefont {Manago}}, \bibinfo
  {author} {\bibfnamefont {Noriyasu}\ \bibnamefont {Chikuchi}}, \bibinfo
  {author} {\bibfnamefont {Hitoshi}\ \bibnamefont {Sugawara}}, \bibinfo
  {author} {\bibfnamefont {Hideki}\ \bibnamefont {Tou}}, \ and\ \bibinfo
  {author} {\bibfnamefont {Hisatomo}\ \bibnamefont {Harima}},\ }\bibfield
  {title} {\enquote {\bibinfo {title} {Indication of ferromagnetic quantum
  critical point in kondo lattice {${\mathrm{CeRh}}_{6}{\mathrm{Ge}}_{4}$}},}\
  }\href {\doibase 10.7566/JPSJ.88.093702} {\bibfield  {journal} {\bibinfo
  {journal} {Journal of the Physical Society of Japan}\ }\textbf {\bibinfo
  {volume} {88}},\ \bibinfo {pages} {093702} (\bibinfo {year}
  {2019})}\BibitemShut {NoStop}%
\bibitem [{\citenamefont {Belitz}\ \emph {et~al.}(1999)\citenamefont {Belitz},
  \citenamefont {Kirkpatrick},\ and\ \citenamefont {Vojta}}]{Belitz_1999}%
  \BibitemOpen
  \bibfield  {author} {\bibinfo {author} {\bibfnamefont {D.}~\bibnamefont
  {Belitz}}, \bibinfo {author} {\bibfnamefont {T.~R.}\ \bibnamefont
  {Kirkpatrick}}, \ and\ \bibinfo {author} {\bibfnamefont {Thomas}\
  \bibnamefont {Vojta}},\ }\bibfield  {title} {\enquote {\bibinfo {title}
  {First order transitions and multicritical points in weak itinerant
  ferromagnets},}\ }\href {\doibase 10.1103/PhysRevLett.82.4707} {\bibfield
  {journal} {\bibinfo  {journal} {Phys. Rev. Lett.}\ }\textbf {\bibinfo
  {volume} {82}},\ \bibinfo {pages} {4707--4710} (\bibinfo {year}
  {1999})}\BibitemShut {NoStop}%
\bibitem [{\citenamefont {Chubukov}\ \emph {et~al.}(2004)\citenamefont
  {Chubukov}, \citenamefont {P\'epin},\ and\ \citenamefont
  {Rech}}]{Chubukov_2004}%
  \BibitemOpen
  \bibfield  {author} {\bibinfo {author} {\bibfnamefont {Andrey~V.}\
  \bibnamefont {Chubukov}}, \bibinfo {author} {\bibfnamefont {Catherine}\
  \bibnamefont {P\'epin}}, \ and\ \bibinfo {author} {\bibfnamefont {Jerome}\
  \bibnamefont {Rech}},\ }\bibfield  {title} {\enquote {\bibinfo {title}
  {Instability of the quantum-critical point of itinerant ferromagnets},}\
  }\href {\doibase 10.1103/PhysRevLett.92.147003} {\bibfield  {journal}
  {\bibinfo  {journal} {Phys. Rev. Lett.}\ }\textbf {\bibinfo {volume} {92}},\
  \bibinfo {pages} {147003} (\bibinfo {year} {2004})}\BibitemShut {NoStop}%
\bibitem [{\citenamefont {Wang}\ \emph {et~al.}(2021)\citenamefont {Wang},
  \citenamefont {Du}, \citenamefont {Zhang}, \citenamefont {Graf},
  \citenamefont {Shen}, \citenamefont {Chen}, \citenamefont {Liu},
  \citenamefont {Smidman}, \citenamefont {Cao}, \citenamefont {Steglich},\ and\
  \citenamefont {Yuan}}]{Wang_2021}%
  \BibitemOpen
  \bibfield  {author} {\bibinfo {author} {\bibfnamefont {An}~\bibnamefont
  {Wang}}, \bibinfo {author} {\bibfnamefont {Feng}\ \bibnamefont {Du}},
  \bibinfo {author} {\bibfnamefont {Yongjun}\ \bibnamefont {Zhang}}, \bibinfo
  {author} {\bibfnamefont {David}\ \bibnamefont {Graf}}, \bibinfo {author}
  {\bibfnamefont {Bin}\ \bibnamefont {Shen}}, \bibinfo {author} {\bibfnamefont
  {Ye}~\bibnamefont {Chen}}, \bibinfo {author} {\bibfnamefont {Yang}\
  \bibnamefont {Liu}}, \bibinfo {author} {\bibfnamefont {Michael}\ \bibnamefont
  {Smidman}}, \bibinfo {author} {\bibfnamefont {Chao}\ \bibnamefont {Cao}},
  \bibinfo {author} {\bibfnamefont {Frank}\ \bibnamefont {Steglich}}, \ and\
  \bibinfo {author} {\bibfnamefont {Huiqiu}\ \bibnamefont {Yuan}},\ }\bibfield
  {title} {\enquote {\bibinfo {title} {Localized 4f-electrons in the quantum
  critical heavy-fermion ferromagnet
  {$\mathrm{Ce}{\mathrm{Rh}}_{6}{\mathrm{Ge}}_{4}$}},}\ }\href {\doibase
  https://doi.org/10.1016/j.scib.2021.03.006} {\bibfield  {journal} {\bibinfo
  {journal} {Science Bulletin}\ }\textbf {\bibinfo {volume} {66}},\ \bibinfo
  {pages} {1389} (\bibinfo {year} {2021})}\BibitemShut {NoStop}%
\bibitem [{\citenamefont {Wu}\ \emph {et~al.}(2021)\citenamefont {Wu},
  \citenamefont {Zhang}, \citenamefont {Du}, \citenamefont {Shen},
  \citenamefont {Zheng}, \citenamefont {Fang}, \citenamefont {Smidman},
  \citenamefont {Cao}, \citenamefont {Steglich}, \citenamefont {Yuan},
  \citenamefont {Denlinger},\ and\ \citenamefont {Liu}}]{Yi_2021}%
  \BibitemOpen
  \bibfield  {author} {\bibinfo {author} {\bibfnamefont {Yi}~\bibnamefont
  {Wu}}, \bibinfo {author} {\bibfnamefont {Yongjun}\ \bibnamefont {Zhang}},
  \bibinfo {author} {\bibfnamefont {Feng}\ \bibnamefont {Du}}, \bibinfo
  {author} {\bibfnamefont {Bin}\ \bibnamefont {Shen}}, \bibinfo {author}
  {\bibfnamefont {Hao}\ \bibnamefont {Zheng}}, \bibinfo {author} {\bibfnamefont
  {Yuan}\ \bibnamefont {Fang}}, \bibinfo {author} {\bibfnamefont {Michael}\
  \bibnamefont {Smidman}}, \bibinfo {author} {\bibfnamefont {Chao}\
  \bibnamefont {Cao}}, \bibinfo {author} {\bibfnamefont {Frank}\ \bibnamefont
  {Steglich}}, \bibinfo {author} {\bibfnamefont {Huiqiu}\ \bibnamefont {Yuan}},
  \bibinfo {author} {\bibfnamefont {Jonathan~D.}\ \bibnamefont {Denlinger}}, \
  and\ \bibinfo {author} {\bibfnamefont {Yang}\ \bibnamefont {Liu}},\
  }\bibfield  {title} {\enquote {\bibinfo {title} {Anisotropic
  $c\ensuremath{-}f$ hybridization in the ferromagnetic quantum critical metal
  {${\mathrm{CeRh}}_{6}{\mathrm{Ge}}_{4}$}},}\ }\href {\doibase
  10.1103/PhysRevLett.126.216406} {\bibfield  {journal} {\bibinfo  {journal}
  {Phys. Rev. Lett.}\ }\textbf {\bibinfo {volume} {126}},\ \bibinfo {pages}
  {216406} (\bibinfo {year} {2021})}\BibitemShut {NoStop}%
\bibitem [{\citenamefont {Pei}\ \emph {et~al.}(2021)\citenamefont {Pei},
  \citenamefont {Zhang}, \citenamefont {Wei}, \citenamefont {Chen},
  \citenamefont {Hu}, \citenamefont {Yang}, \citenamefont {Yuan},\ and\
  \citenamefont {Qi}}]{Pei_2021}%
  \BibitemOpen
  \bibfield  {author} {\bibinfo {author} {\bibfnamefont {Y.~H.}\ \bibnamefont
  {Pei}}, \bibinfo {author} {\bibfnamefont {Y.~J.}\ \bibnamefont {Zhang}},
  \bibinfo {author} {\bibfnamefont {Z.~X.}\ \bibnamefont {Wei}}, \bibinfo
  {author} {\bibfnamefont {Y.~X.}\ \bibnamefont {Chen}}, \bibinfo {author}
  {\bibfnamefont {K.}~\bibnamefont {Hu}}, \bibinfo {author} {\bibfnamefont
  {Yi-feng}\ \bibnamefont {Yang}}, \bibinfo {author} {\bibfnamefont {H.~Q.}\
  \bibnamefont {Yuan}}, \ and\ \bibinfo {author} {\bibfnamefont
  {J.}~\bibnamefont {Qi}},\ }\bibfield  {title} {\enquote {\bibinfo {title}
  {Unveiling the hybridization process in a quantum critical ferromagnet by
  ultrafast optical spectroscopy},}\ }\href {\doibase
  10.1103/PhysRevB.103.L180409} {\bibfield  {journal} {\bibinfo  {journal}
  {Phys. Rev. B}\ }\textbf {\bibinfo {volume} {103}},\ \bibinfo {pages}
  {L180409} (\bibinfo {year} {2021})}\BibitemShut {NoStop}%
\bibitem [{\citenamefont {Shu}\ \emph {et~al.}(2021)\citenamefont {Shu},
  \citenamefont {Adroja}, \citenamefont {Hillier}, \citenamefont {Zhang},
  \citenamefont {Chen}, \citenamefont {Shen}, \citenamefont {Orlandi},
  \citenamefont {Walker}, \citenamefont {Liu}, \citenamefont {Cao},
  \citenamefont {Steglich}, \citenamefont {Yuan},\ and\ \citenamefont
  {Smidman}}]{Shu_2021}%
  \BibitemOpen
  \bibfield  {author} {\bibinfo {author} {\bibfnamefont {J.~W.}\ \bibnamefont
  {Shu}}, \bibinfo {author} {\bibfnamefont {D.~T.}\ \bibnamefont {Adroja}},
  \bibinfo {author} {\bibfnamefont {A.~D.}\ \bibnamefont {Hillier}}, \bibinfo
  {author} {\bibfnamefont {Y.~J.}\ \bibnamefont {Zhang}}, \bibinfo {author}
  {\bibfnamefont {Y.~X.}\ \bibnamefont {Chen}}, \bibinfo {author}
  {\bibfnamefont {B.}~\bibnamefont {Shen}}, \bibinfo {author} {\bibfnamefont
  {F.}~\bibnamefont {Orlandi}}, \bibinfo {author} {\bibfnamefont {H.~C.}\
  \bibnamefont {Walker}}, \bibinfo {author} {\bibfnamefont {Y.}~\bibnamefont
  {Liu}}, \bibinfo {author} {\bibfnamefont {C.}~\bibnamefont {Cao}}, \bibinfo
  {author} {\bibfnamefont {F.}~\bibnamefont {Steglich}}, \bibinfo {author}
  {\bibfnamefont {H.~Q.}\ \bibnamefont {Yuan}}, \ and\ \bibinfo {author}
  {\bibfnamefont {M.}~\bibnamefont {Smidman}},\ }\bibfield  {title} {\enquote
  {\bibinfo {title} {Magnetic order and crystalline electric field excitations
  of the quantum critical heavy-fermion ferromagnet
  {$\mathrm{Ce}{\mathrm{Rh}}_{6}{\mathrm{Ge}}_{4}$}},}\ }\href {\doibase
  10.1103/PhysRevB.104.L140411} {\bibfield  {journal} {\bibinfo  {journal}
  {Phys. Rev. B}\ }\textbf {\bibinfo {volume} {104}},\ \bibinfo {pages}
  {L140411} (\bibinfo {year} {2021})}\BibitemShut {NoStop}%
\bibitem [{\citenamefont {Zhu}\ \emph {et~al.}(2003)\citenamefont {Zhu},
  \citenamefont {Garst}, \citenamefont {Rosch},\ and\ \citenamefont
  {Si}}]{Zhu_2003}%
  \BibitemOpen
  \bibfield  {author} {\bibinfo {author} {\bibfnamefont {Lijun}\ \bibnamefont
  {Zhu}}, \bibinfo {author} {\bibfnamefont {Markus}\ \bibnamefont {Garst}},
  \bibinfo {author} {\bibfnamefont {Achim}\ \bibnamefont {Rosch}}, \ and\
  \bibinfo {author} {\bibfnamefont {Qimiao}\ \bibnamefont {Si}},\ }\bibfield
  {title} {\enquote {\bibinfo {title} {Universally diverging gr\"uneisen
  parameter and the magnetocaloric effect close to quantum critical points},}\
  }\href {\doibase 10.1103/PhysRevLett.91.066404} {\bibfield  {journal}
  {\bibinfo  {journal} {Phys. Rev. Lett.}\ }\textbf {\bibinfo {volume} {91}},\
  \bibinfo {pages} {066404} (\bibinfo {year} {2003})}\BibitemShut {NoStop}%
\bibitem [{\citenamefont {Steppke}\ \emph {et~al.}(2013)\citenamefont
  {Steppke}, \citenamefont {K{\"u}chler}, \citenamefont {Lausberg},
  \citenamefont {Lengyel}, \citenamefont {Steinke}, \citenamefont {Borth},
  \citenamefont {L{\"u}hmann}, \citenamefont {Krellner}, \citenamefont
  {Nicklas}, \citenamefont {Geibel}, \citenamefont {Steglich},\ and\
  \citenamefont {Brando}}]{Steppke_2013}%
  \BibitemOpen
  \bibfield  {author} {\bibinfo {author} {\bibfnamefont {Alexander}\
  \bibnamefont {Steppke}}, \bibinfo {author} {\bibfnamefont {Robert}\
  \bibnamefont {K{\"u}chler}}, \bibinfo {author} {\bibfnamefont {Stefan}\
  \bibnamefont {Lausberg}}, \bibinfo {author} {\bibfnamefont {Edit}\
  \bibnamefont {Lengyel}}, \bibinfo {author} {\bibfnamefont {Lucia}\
  \bibnamefont {Steinke}}, \bibinfo {author} {\bibfnamefont {Robert}\
  \bibnamefont {Borth}}, \bibinfo {author} {\bibfnamefont {Thomas}\
  \bibnamefont {L{\"u}hmann}}, \bibinfo {author} {\bibfnamefont {Cornelius}\
  \bibnamefont {Krellner}}, \bibinfo {author} {\bibfnamefont {Michael}\
  \bibnamefont {Nicklas}}, \bibinfo {author} {\bibfnamefont {Christoph}\
  \bibnamefont {Geibel}}, \bibinfo {author} {\bibfnamefont {Frank}\
  \bibnamefont {Steglich}}, \ and\ \bibinfo {author} {\bibfnamefont {Manuel}\
  \bibnamefont {Brando}},\ }\bibfield  {title} {\enquote {\bibinfo {title}
  {Ferromagnetic quantum critical point in the heavy-fermion metal
  {${\mathrm{YbNi}}_{4} {\mathrm{
  ({\mathrm{P}}_{1\ensuremath{-}x}{\mathrm{As}}_{x}) }}_{2}$}},}\ }\href
  {\doibase 10.1126/science.1230583} {\bibfield  {journal} {\bibinfo  {journal}
  {Science}\ }\textbf {\bibinfo {volume} {339}},\ \bibinfo {pages} {933--936}
  (\bibinfo {year} {2013})}\BibitemShut {NoStop}%
\bibitem [{\citenamefont {Schr\"{o}der}\ \emph {et~al.}(2000)\citenamefont
  {Schr\"{o}der}, \citenamefont {Aeppli}, \citenamefont {Adams}, \citenamefont
  {Stockert}, \citenamefont {L\"{o}hneysen}, \citenamefont {Bucher},
  \citenamefont {Ramazashvili},\ and\ \citenamefont {Coleman}}]{CeCu6_2000}%
  \BibitemOpen
  \bibfield  {author} {\bibinfo {author} {\bibfnamefont {A.}~\bibnamefont
  {Schr\"{o}der}}, \bibinfo {author} {\bibfnamefont {G.}~\bibnamefont
  {Aeppli}}, \bibinfo {author} {\bibfnamefont {M.}~\bibnamefont {Adams}},
  \bibinfo {author} {\bibfnamefont {O.}~\bibnamefont {Stockert}}, \bibinfo
  {author} {\bibfnamefont {H.v.}\ \bibnamefont {L\"{o}hneysen}}, \bibinfo
  {author} {\bibfnamefont {E.}~\bibnamefont {Bucher}}, \bibinfo {author}
  {\bibfnamefont {R.}~\bibnamefont {Ramazashvili}}, \ and\ \bibinfo {author}
  {\bibfnamefont {P.}~\bibnamefont {Coleman}},\ }\bibfield  {title} {\enquote
  {\bibinfo {title} {Onset of antiferromagnetism in heavy-fermion metals},}\
  }\href {\doibase 10.1038/35030039} {\bibfield  {journal} {\bibinfo  {journal}
  {Nature}\ }\textbf {\bibinfo {volume} {407}},\ \bibinfo {pages} {351--355}
  (\bibinfo {year} {2000})}\BibitemShut {NoStop}%
\bibitem [{\citenamefont {Matsuoka}\ \emph {et~al.}(2015)\citenamefont
  {Matsuoka}, \citenamefont {Hondo}, \citenamefont {Fujii}, \citenamefont
  {Oshima}, \citenamefont {Sugawara}, \citenamefont {Sakurai}, \citenamefont
  {Ohta}, \citenamefont {Kneidinger}, \citenamefont {Salamakha}, \citenamefont
  {Michor},\ and\ \citenamefont {Bauer}}]{Matsuoka_2015}%
  \BibitemOpen
  \bibfield  {author} {\bibinfo {author} {\bibfnamefont {Eiichi}\ \bibnamefont
  {Matsuoka}}, \bibinfo {author} {\bibfnamefont {Chisato}\ \bibnamefont
  {Hondo}}, \bibinfo {author} {\bibfnamefont {Tatsuya}\ \bibnamefont {Fujii}},
  \bibinfo {author} {\bibfnamefont {Akihiro}\ \bibnamefont {Oshima}}, \bibinfo
  {author} {\bibfnamefont {Hitoshi}\ \bibnamefont {Sugawara}}, \bibinfo
  {author} {\bibfnamefont {Takahiro}\ \bibnamefont {Sakurai}}, \bibinfo
  {author} {\bibfnamefont {Hitoshi}\ \bibnamefont {Ohta}}, \bibinfo {author}
  {\bibfnamefont {Friedrich}\ \bibnamefont {Kneidinger}}, \bibinfo {author}
  {\bibfnamefont {Leonid}\ \bibnamefont {Salamakha}}, \bibinfo {author}
  {\bibfnamefont {Herwig}\ \bibnamefont {Michor}}, \ and\ \bibinfo {author}
  {\bibfnamefont {Ernst}\ \bibnamefont {Bauer}},\ }\bibfield  {title} {\enquote
  {\bibinfo {title} {Ferromagnetic transition at 2.5 {K} in the hexagonal
  kondo-lattice compound {${\mathrm{CeRh}}_{6}{\mathrm{Ge}}_{4}$}},}\ }\href
  {\doibase 10.7566/JPSJ.84.073704} {\bibfield  {journal} {\bibinfo  {journal}
  {Journal of the Physical Society of Japan}\ }\textbf {\bibinfo {volume}
  {84}},\ \bibinfo {pages} {073704} (\bibinfo {year} {2015})}\BibitemShut
  {NoStop}%
\bibitem [{\citenamefont {Vo{\ss}winkel}\ \emph {et~al.}(2012)\citenamefont
  {Vo{\ss}winkel}, \citenamefont {Niehaus}, \citenamefont {Rodewald},\ and\
  \citenamefont {P\"{o}ttgen}}]{Daniel_2012}%
  \BibitemOpen
  \bibfield  {author} {\bibinfo {author} {\bibfnamefont {D.}~\bibnamefont
  {Vo{\ss}winkel}}, \bibinfo {author} {\bibfnamefont {O.}~\bibnamefont
  {Niehaus}}, \bibinfo {author} {\bibfnamefont {U.~C.}\ \bibnamefont
  {Rodewald}}, \ and\ \bibinfo {author} {\bibfnamefont {R.}~\bibnamefont
  {P\"{o}ttgen}},\ }\bibfield  {title} {\enquote {\bibinfo {title} {Bismuth
  flux growth of {${\mathrm{CeRh}}_{6}{\mathrm{Ge}}_{4}$} and
  {${\mathrm{CeRh}}_{2}{\mathrm{Ge}}_{2}$} single crystals},}\ }\href {\doibase
  10.5560/znb.2012-0265} {\bibfield  {journal} {\bibinfo  {journal}
  {Zeitschrift f\"{u}r Naturforschung B}\ }\textbf {\bibinfo {volume} {67}},\
  \bibinfo {pages} {1241--1247} (\bibinfo {year} {2012})}\BibitemShut {NoStop}%
\bibitem [{\citenamefont {Birch}(1947)}]{Birch_1947}%
  \BibitemOpen
  \bibfield  {author} {\bibinfo {author} {\bibfnamefont {Francis}\ \bibnamefont
  {Birch}},\ }\bibfield  {title} {\enquote {\bibinfo {title} {Finite elastic
  strain of cubic crystals},}\ }\href {\doibase 10.1103/PhysRev.71.809}
  {\bibfield  {journal} {\bibinfo  {journal} {Phys. Rev.}\ }\textbf {\bibinfo
  {volume} {71}},\ \bibinfo {pages} {809--824} (\bibinfo {year}
  {1947})}\BibitemShut {NoStop}%
\bibitem [{\citenamefont {Xu}\ \emph {et~al.}(2021)\citenamefont {Xu},
  \citenamefont {Su}, \citenamefont {Kumar}, \citenamefont {Luo}, \citenamefont
  {Nie}, \citenamefont {Wang}, \citenamefont {Du}, \citenamefont {Li},
  \citenamefont {Smidman},\ and\ \citenamefont {Yuan}}]{Cheng_2021}%
  \BibitemOpen
  \bibfield  {author} {\bibinfo {author} {\bibfnamefont {Jia-Cheng}\
  \bibnamefont {Xu}}, \bibinfo {author} {\bibfnamefont {Hang}\ \bibnamefont
  {Su}}, \bibinfo {author} {\bibfnamefont {Rohit}\ \bibnamefont {Kumar}},
  \bibinfo {author} {\bibfnamefont {Shuai-Shuai}\ \bibnamefont {Luo}}, \bibinfo
  {author} {\bibfnamefont {Zhi-Yong}\ \bibnamefont {Nie}}, \bibinfo {author}
  {\bibfnamefont {An}~\bibnamefont {Wang}}, \bibinfo {author} {\bibfnamefont
  {Feng}\ \bibnamefont {Du}}, \bibinfo {author} {\bibfnamefont {Rui}\
  \bibnamefont {Li}}, \bibinfo {author} {\bibfnamefont {Michael}\ \bibnamefont
  {Smidman}}, \ and\ \bibinfo {author} {\bibfnamefont {Hui-Qiu}\ \bibnamefont
  {Yuan}},\ }\bibfield  {title} {\enquote {\bibinfo {title} {Ce-site dilution
  in the ferromagnetic kondo lattice
  {${\mathrm{CeRh}}_{6}{\mathrm{Ge}}_{4}$}},}\ }\href {\doibase
  10.1088/0256-307x/38/8/087101} {\bibfield  {journal} {\bibinfo  {journal}
  {Chinese Physics Letters}\ }\textbf {\bibinfo {volume} {38}},\ \bibinfo
  {pages} {087101} (\bibinfo {year} {2021})}\BibitemShut {NoStop}%
\bibitem [{\citenamefont {Rosch}(1999)}]{Rosch_1999}%
  \BibitemOpen
  \bibfield  {author} {\bibinfo {author} {\bibfnamefont {A.}~\bibnamefont
  {Rosch}},\ }\bibfield  {title} {\enquote {\bibinfo {title} {Interplay of
  disorder and spin fluctuations in the resistivity near a quantum critical
  point},}\ }\href {\doibase 10.1103/PhysRevLett.82.4280} {\bibfield  {journal}
  {\bibinfo  {journal} {Phys. Rev. Lett.}\ }\textbf {\bibinfo {volume} {82}},\
  \bibinfo {pages} {4280--4283} (\bibinfo {year} {1999})}\BibitemShut {NoStop}%
\bibitem [{\citenamefont {L\"ohneysen}\ \emph {et~al.}(2007)\citenamefont
  {L\"ohneysen}, \citenamefont {Rosch}, \citenamefont {Vojta},\ and\
  \citenamefont {W\"olfle}}]{Hilbert_2007}%
  \BibitemOpen
  \bibfield  {author} {\bibinfo {author} {\bibfnamefont {Hilbert~v.}\
  \bibnamefont {L\"ohneysen}}, \bibinfo {author} {\bibfnamefont {Achim}\
  \bibnamefont {Rosch}}, \bibinfo {author} {\bibfnamefont {Matthias}\
  \bibnamefont {Vojta}}, \ and\ \bibinfo {author} {\bibfnamefont {Peter}\
  \bibnamefont {W\"olfle}},\ }\bibfield  {title} {\enquote {\bibinfo {title}
  {Fermi-liquid instabilities at magnetic quantum phase transitions},}\ }\href
  {\doibase 10.1103/RevModPhys.79.1015} {\bibfield  {journal} {\bibinfo
  {journal} {Rev. Mod. Phys.}\ }\textbf {\bibinfo {volume} {79}},\ \bibinfo
  {pages} {1015--1075} (\bibinfo {year} {2007})}\BibitemShut {NoStop}%
\end{thebibliography}
\end{document}